\documentclass[aps,prb,preprint,floatfix,showpacs]{revtex4-1}
\usepackage[]{graphicx}
\usepackage{epstopdf}
\usepackage{bm}
\usepackage{framed}
\newcommand{\eg}{e.g.,}
\newcommand{\ie}{i.e.,}
\begin{document}
\title{Distinguishing quantum dot-like localized states from quantum well-like extended states across the exciton emission line in a quantum well}

\author{Sumi Bhuyan}
\affiliation{Indian Institute of Science Education and Research Kolkata, Mohanpur, Nadia 741246, West Bengal, India}

\author{Richarj Mondal}
\affiliation{Indian Institute of Science Education and Research Kolkata, Mohanpur, Nadia 741246, West Bengal, India}

\author{Bipul Pal}\email{bipul@iiserkol.ac.in}
\affiliation{Indian Institute of Science Education and Research Kolkata, Mohanpur, Nadia 741246, West Bengal, India}

\author{Bhavtosh Bansal}\email{bhavtosh@iiserkol.ac.in}
\affiliation{Indian Institute of Science Education and Research Kolkata, Mohanpur, Nadia 741246, West Bengal, India}

\date{\today}
\begin{abstract}
We have closely examined the emission spectrum at the heavy-hole exciton resonance in a high-quality GaAs multi-quantum well (MQW) sample using picosecond excitation-correlation photoluminescence (ECPL) spectroscopy. Dynamics of the ECPL signal at low and high energy sides of the excitonic photoluminescence (PL) peak shows complementary behavior. The ECPL signal is positive (negative) below (above) the PL peak and it changes sign within a narrow band of energy lying between excitonic absorption and emission peaks. The energy at which this sign change takes place is interpreted as the excitonic mobility edge as it separates localized excitons in quantum dot-like states from mobile excitons in quantum well-like states. 

\end{abstract}
\maketitle

\section{Introduction}
While the operating wavelength of light emitters and detectors is primarily determined by the energy position of the density of states, the nature of these states themselves---whether they are localized or extended---has a profound effect on the device characteristics. While localization may reduce the responsivity in detectors, it can also result in many useful attributes:~\cite{Kavokin} giant oscillator strength,~\cite{Pelant-Valenta, naeem} weak temperature dependence of optical properties,~\cite{ODonnell} strong electron-phonon coupling,~\cite{Alkauskas} which in turn can lead to a transition to strong exciton-photon coupling,~\cite{Andreani} and persistence of luminescence up to high temperatures.~\cite{ODonnell} But if carrier localization is to be looked upon as an engineering parameter, techniques are needed to reliably characterize these localized states.

Focusing on a well understood and very clean system, GaAs quantum wells (QWs),~\cite{runge, Hegarty, Herman-Bimberg-Christen} in this paper we have carried out time-resolved measurements to understand the spectral diffusion of carriers across the emission line and  determine the location of the mobility edge (defined below). The technique is directly applicable to any other semiconductor.

The problem of interaction of the exciton with disorder is non-trivial because the Wannier exciton must be treated as an irreducible two-particle (electron-hole) bound state of a finite (non-zero) size.~\cite{runge} The disorder potential affects the center-of-mass (CoM) degree of freedom, which is quiescent in most of the other optical phenomena. Strong localization of the CoM of the exciton effectively implies a transition to quantum dot (QD)-like states.~\cite{Adams}

Consider the excitonic emission or absorption spectrum resulting from the excitation of a macroscopic area of a sample. In completely disorder-free systems, the electron and hole states which comprise the exciton are well-described by the Bloch wavefunctions. One would thus have a band of lifetime-broadened (hydrogenic) exciton states labeled by the CoM wave vector $\vec{K}_{CM}$ with the ground state energy dispersion $E(K_{CM})=E_g-E_B+{\hbar^2\over 2M}K_{CM}^2$, where $E_g$ is the band-gap energy, $E_B$ is the exciton binding energy and $M$ is the exciton mass.  Since only $K_{CM}\approx 0$ excitons couple to light, one should observe a single homogeneously broadened excitonic peak in the optical density of states. This peak should arise at energy $E=E_g-E_B$. The homogeneous linewidth $\Gamma$ is bounded from below by the inverse of the radiative lifetime of a few nanosecond though other scattering process may substantially enhance it to a few picoseconds inverse  (few meV) depending on temperature.

\begin{figure}[!b]
	\includegraphics[clip,width=7.0cm]{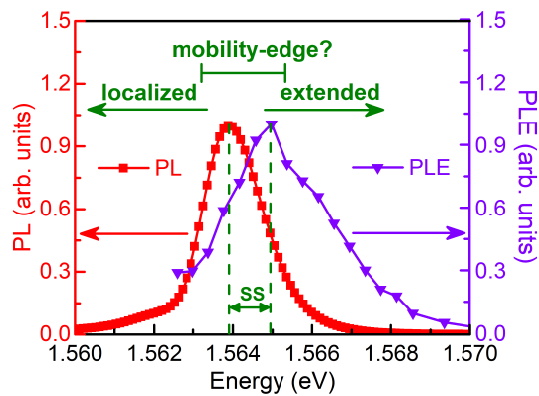}
	\caption{Photoluminescence (PL) and photoluminescence excitation (PLE) spectra due to 1s heavy-hole excitons from 8~nm GaAs QWs at 4~K. The Stokes shift (SS) as well as the energy positions of the localized and extended exciton states are schematically marked. One would like to determine the mobility edge separating these two kinds of states.~\cite{Mott_Argument}}  \label{PL-PLE}
\end{figure}
\begin{figure*}[!t]
	\includegraphics[clip,width=15cm]{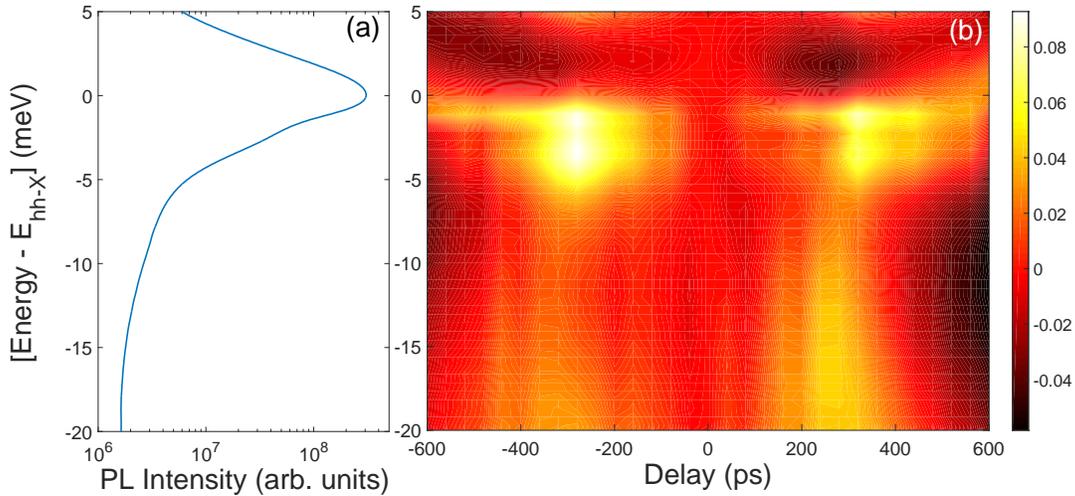}
	\caption{(a) PL spectrum of Fig.~\ref{PL-PLE} on a semilogarithmic scale. (b) Spectrally- and temporally-resolved ECPL signal is shown as a two-dimensional color plot. In both the graphs, the zero of the energy scale is shifted to coincide with the 1s heavy-hole exciton (hh-x) PL peak.}  \label{ECPL-2D}
\end{figure*}

Realistically, we must consider a system with additional static disorder -- interface roughness, site defects due to alloying and doping, and dislocations -- leading to further inhomogeneous broadening. Depending on the strength of the perturbation caused by this disorder, one may invoke the following two regimes:

{\em (i) Weak disorder regime with extended states.} $\vec{K}_{CM}$ may still be used to label the states but the optical selection rule $K_{CM}\approx 0$ may not be completely valid due to the extra momentum being taken by the disorder potential. Emission would thus be inhomogeneously broadened (sum of states with different $K_{CM}$) resulting from emission from states with $K_{CM}\neq 0$. These states are extended Bloch states.

{\em (ii) Strong disorder regime with localized states.} The other extreme is that of very strong disorder which breaks ergodicity of the system; excitons are localized in the QD-like potential wells and the sample behaves like an ensemble of these localized emission and absorption centers. Thus the label $\vec{K}_{CM}$ (that follows from the translational invariance) loses its meaning. Thermal activation of carriers being an inelastic process, does not help to preserve the meaning of $\vec{K}_{CM}$. This QD-like energetically-narrow spatially-localized emission has of course long been observed in micro-PL measurements from even good-quality GaAs quantum wells, especially on the low energy side of the PL  spectrum.~\cite{Zrenner}

The question that interests us here is exactly how do these two types of states coexist within the same sample and we would like to determine the location of the `mobility edge', viz. the value of energy separating localized states from the extended states. Following the debate first started by Mott in the context of transport in disordered semiconductors,~\cite{Mott_Argument}  this question has been extensively investigated both theoretically and experimentally using a variety of optical spectroscopic techniques including resonant Rayleigh scattering,~\cite{Hegarty, Belitsky} micro-PL,~\cite{Wu-Grober} and cathodoluminescence.~\cite{Jahn} The general consensus is that the mobility edge occurs around the peak of the absorption band.

In this work we use a variant of the picosecond time-resolved excitation-correlation photoluminescence~\cite{richarj} (ECPL) spectroscopy to determine the location of the exciton mobility edge in a high-quality GaAs/AlGaAs  multi-quantum wells (MQWs) sample. This is a new and simple way of experimentally locating the mobility edge. Being a differential measurement technique, somewhat similar to the modulation spectroscopy,~\cite{modulation-spectroscopy} ECPL is very sensitive to the details of the relaxation dynamics of the electron-hole pairs prior to their radiative recombination leading to the PL emission.~\cite{Havm_APL43_1983} For example, spectral diffusion and Pauli-blocking dynamics are expected to affect relaxation dynamics of free and bound excitons differently. Spectral features of the ECPL spectra should be able to reflect these differences. This has enabled ECPL spectra to be used to monitor existence and dynamical evolution of the electron-hole plasma and free and bound excitons under varying intensity of nonresonant excitation.~\cite{Bipul_PRB65}  As the exciton mobility edge separates the spectral regions of extended (free) excitons from that of localized (bound) excitons, it may be easily and precisely identified from the distinct features of ECPL spectra.

\section{Experimental}
We performed picosecond time-resolved ECPL measurements~\cite{richarj} at 4~K on a molecular beam epitaxy grown GaAs MQWs sample, having 20 periods of 8~nm GaAs wells separated by 15~nm Al$_{0.33}$Ga$_{0.67}$As barriers, using $\sim 100$ femtosecond pulses from a Ti:sapphire laser at 760~nm, thus nonresonantly exciting the sample at about 65~meV above the exciton resonance. We define the ECPL signal $EC(\tau, \hbar\omega)$ at a delay $\tau$ and energy $\hbar \omega$ as
\begin{equation}
EC(\tau, \hbar\omega)\equiv \{PL(\tau, \hbar\omega)-PL(0, \hbar\omega)\}/PL(0, \hbar\omega),
\end{equation}
where $PL(\tau, \hbar\omega)$ is the time-integrated PL signal at an energy $\hbar \omega$ due to two-beam excitation with a delay $\tau$ between the corresponding pulses in the two beams and $PL(0, \hbar\omega)$ is the same quantity at $\tau=0$. This way of defining ECPL signal is  different from the usual definition of ECPL signal found in the literature~\cite{Havm_APL43_1983,Bipul_PRB65,Pau-Kuhl} but is simpler to interpret.~\cite{richarj} We measured PL spectra at different delays using high resolution CCD-spectrograph and calculated ECPL spectra numerically from the stored PL spectra. Details of our experimental scheme was described in Ref.~\onlinecite{richarj}. This method, giving high spectral resolution, is different from the conventional ECPL measurements where the two beams incident on the sample are chopped (modulated) at different frequencies and the emitted PL signal (after being spectrally dispersed) is measured with a point detector at the sum or difference frequency, one wavelength at a time.~\cite{Havm_APL43_1983,Bipul_PRB65,Pau-Kuhl}     

As per our definition, ECPL signal is zero at $\tau=0$ and should remain at zero at any other delay if there are no nonlinearities in the PL emission, as the total number of photons incident on the sample per second is independent of the delay between the pulses. A nonzero ECPL signal is indicative of nonlinearity in the emitted PL caused by various nonlinear processes such as phase space filling, band gap renormalization, spectral broadening and spectral diffusion, which determine carrier dynamics in semiconductors. The spectrally-resolved ECPL signal is just the relative fractional change in the number of emitted photons as a function of energy at a given delay with respect to zero delay. A positive (negative) ECPL signal at a nonzero delay at a given energy implies that more (less) photons are emitted per second at this energy when there is a time delay between the arrival of pulses in the two excitation beams compared to the situation when the pulses in the two excitation beams arrive simultaneously at the sample, {\ie} $\tau=0$.

\section{Results and discussions}
Figure~\ref{PL-PLE} shows the steady-state PL and PLE spectra of our sample at 4~K. Single narrow peaks at 1.565 and 1.564~eV, respectively, in PLE and PL spectra are assigned to the ground state (1s) heavy-hole excitonic (hh-x) absorption and emission. As usual, the PL spectrum is Stokes-shifted with respect to the PLE spectrum by about 1~meV, which is a little less than the statistical tomography~\cite{Yang} estimate (Stokes shift $\approx 0.55$ times absorption linewidth) of 1.6~meV for the PLE full-width-at-half-maximum of about 3~meV.  This Stokes shift is understood to be due to the presence of lower energy localized states which are selectively picked out by the PL spectrum as the carriers photoexcited well above the band edge relax into these. On the other hand the PLE spectrum (which approximates an absorption spectrum) will be proportional to the actual optical density of states.~\cite{runge, Pelant-Valenta,PLE} The narrow linewidths of the PL and PLE spectra, and the small Stokes shift between them between them demonstrate the good optical quality of this sample.

Figure~\ref{ECPL-2D}(b) is a two-dimensional (2D) color plot of the ECPL signal in the delay-energy plane around the 1s hh-x PL peak. For comparison, the PL spectrum from Fig.~\ref{PL-PLE} is replotted on a semilogarithmic scale in Fig.~\ref{ECPL-2D}(a). The zero of the energy scale has been shifted to coincide with the 1s hh-x PL peak. Note that the ECPL signal is $\lesssim \pm 10\%$ of the PL signal. Since both the incident beams are of equal fluence, the ECPL signal should theoretically be symmetric with respect to the zero delay. This is roughly verified in Fig.~\ref{ECPL-2D}(b).
\begin{figure}[htb]
	\includegraphics[clip,width=6.5cm]{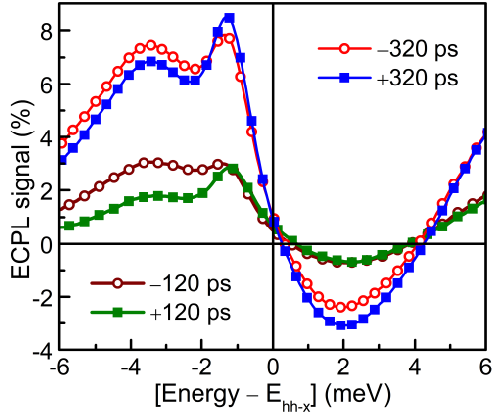}
	\caption{ECPL spectra at a few delays. The zero of the energy scale is shifted to coincide with the 1s heavy-hole exciton PL peak. For visual clarity, eight data points are skipped for every two plotted data points. ECPL signal is mostly negative (positive) at energies above (below) the exciton peak and the zero-crossing of the ECPL signal takes place slightly above the PL peak.}  \label{EC-spectra}
\end{figure}

The ECPL signal in Fig.~\ref{ECPL-2D}(b) is rich in features. Most interesting and relevant (for this work) feature is that the ECPL signal is mostly positive (negative) at the lower (higher) energy side of the 1s hh-x PL peak and it changes sign within a narrow band of energy above the 1s hh-x PL peak. This is more clearly seen in Fig.~\ref{EC-spectra}, where ECPL spectra [vertical slices from Fig.~\ref{ECPL-2D}(b)] at a few different delays ($\pm 120$ and $\pm 320$~ps) are plotted. The exact energy-position of zero-crossing of the ECPL signal is slightly different at different delays (varies within an energy band of $\sim 1$~meV). Parenthetically, also note that in Fig.~\ref{ECPL-2D}(b), a cleanly measurable ECPL signal is observed even at 20~meV below the exciton peak where the PL signal is nearly zero. This may arise from various low-energy defect states present with very low density of states within the GaAs bandgap. Thus ECPL spectroscopy may be used to study properties of defect states which are not well accessible in PL measurements. For example, a bound exciton feature, hardly visible in PL spectrum, is also observed at $\approx 3$~meV below the 1s hh-x PL peak.

To understand the physical processes underlying the observations, note that within the first few tens of picoseconds after the nonresonant excitation ($\approx 65$~meV detuning) above the exciton absorption peak, the carriers will `roll down' to progressively lower energy states, emitting acoustic phonons, until they reach close to the bottom of the respective bands and bind to form excitons. If they encounter a localized state of energy barrier sufficiently larger than the thermal energy, one can assume that the excitons essentially get trapped on first such encounter,~\cite{Yang} until they finally annihilate radiatively or nonradiatively or go out of the trap due to thermal activation. It is also possible that the excitons do not encounter any localized state (or all such localized states they come close to are already filled). In that case the CoM exciton wave function would remain extended. Due to the higher binding energy, the trapped exciton levels will be at a lower energy compared to the extended states. 

Depending upon the temperature of the experiment, thermally activated hopping of carriers out of this trapping sites can take place within a timescale comparable to the radiative lifetime of carriers. At any finite temperature, carrier localization will have a finite lifetime, $\tau_a = \nu^{-1} \exp(\Delta E/k_B T)$, where $\Delta E$ is the trapping energy, $k_B$ is the Boltzmann constant, $T$ is the temperature, and $\nu$ is the attempt frequency. If we take $\nu \sim 10^{12}$~s$^{-1}$ (typical order of magnitude of phonon frequency in GaAs),  $k_B T=0.34$~meV, corresponding to our experimental temperature of 4~K, and $\tau_a \sim 100$~ps (within our measured delay range and less than the radiative lifetime of excitons in high quality GaAs QWs), we get $\Delta E \approx 1.6$~meV. So, activated hopping of localized carriers may take place in our experiment within an energy range of about 1.6~meV. Of course, this is a rough estimate because both the phonon frequency and the carrier recombination times cannot be given a single definite value and temperature only adds an exponential weight to the inherently probabilistic hopping processes. 

The positive sign of the ECPL signal below the 1s hh-x PL peak [Figs.~\ref{ECPL-2D}(b) and \ref{EC-spectra}] is indicative of more photons being recorded in the PL at that energy when the excitation pulses in the two beams are incident on the sample with a nonzero delay as compared to when they are temporally coincident. The opposite trend is observed at energies  above the exciton peak, with the ECPL signal going to zero within a narrow energy band slightly above the exciton PL peak. This suggests that the dynamics of photogenerated carriers is different at energies above and below the 1s hh-x PL peak.
\begin{figure}[htb]
    \includegraphics[clip,width=6.5cm]{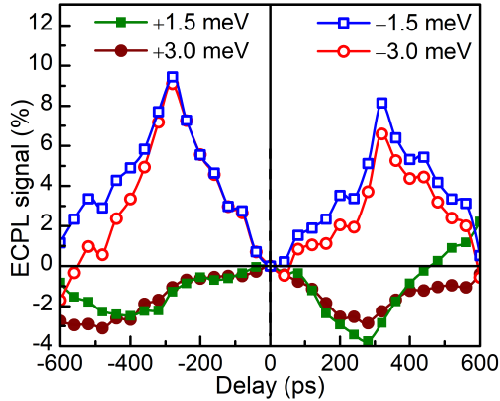}
    \caption{Delay dependence of ECPL signal at a few energies. Opposite but complimentary behavior is seen due to the different dynamics of carriers at energies above and below the 1s hh-x PL peak.}  \label{EC-delay}
\end{figure}

In order to verify this, we plot ECPL signal as a function of delay [horizontal slices from Fig.~\ref{ECPL-2D}(b)] at a few energies above and below the 1s hh-x PL peak [Fig.~\ref{EC-delay}]. Clearly different delay-dependence of ECPL signal at energies above and below the 1s hh-x PL peak is observed. The signal is mostly positive (negative) and increases (decreases) with delay to reach a maximum (minimum) at $\approx 300$~ps and then the signal decreases (increases) at longer delays. This may be understood in the following way. After the nonresonant photoexcitation of free electron-hole pairs and formation of excitons within a few tens of picoseconds, the free and mobile excitons from QW-like flat regions slowly migrate to lower energy localized states in QD-like deeper traps before they radiatively recombine to emit PL. As the higher energy QW-like states feed the lower energy QD-like states with carriers, the ECPL signal is negative for the higher energy states and it is positive for the lower energy states. The localized states will also have substantially larger oscillator strength~\cite{Pelant-Valenta} that would further help them to have a positive ECPL signal. For shorter delays ($\tau \ll $~exciton radiative lifetime $\tau_r$), the second excitation pulse pump in a large number of carriers into the QWs before much of the carriers excited by the first pulse could radiatively decay. As a result, the lower energy localized states with smaller density of states get saturated by carriers and prevent further spectral diffusion of higher energy excitons to these lower energy localized states due to Pauli's exclusion principle. This results in a small positive (negative) signal at energies below (above) the exciton peak. With increasing delay, more of the localized carriers coming from photoexcitation by the first pulse can radiatively decay before the generation of carriers by the delayed second pulse. These states can also be emptied out via thermally activated hopping of carriers out of these trapping sites to further lower energy states. This allows more of the higher energy excitons generated by the second pulse to diffuse to lower energy localized states and radiatively decay from these lower energy states. Due to this delayed diffusion of carriers from higher to lower energy, the positive (negative) signal at energies below (above) exciton peak increases (decreases). This continues until the delay becomes comparable to $\tau_r$. For $\tau > \tau_r$, the Pauli blocking effect and delayed migration of higher energy exciton to lower energy states start reducing. As a result, the positive (negative) signal at energies below (above) exciton PL peak begins to decrease (increase). The ECPL signals in Fig.~\ref{EC-delay} at different energies above and below the 1s hh-x PL peak reach their respective extrema around $\tau \approx 300$~ps. This is comparable to the radiative lifetime of excitons at low temperatures in good-quality GaAs QWs.~\cite{lifetime}

According to the argument made above, it seems reasonable to assume that spectral diffusion from higher to lower energy is favorable for extended QW-like states; if these were QD-like localized states then the carriers once trapped would not spectrally diffuse. Thus the energy position of the sign reversal of the ECPL signal may be  inferred as the mobility edge. The exact energy position of zero-crossing of ECPL signal in our experiment is found to be delay-dependent. The energy variation $\sim 1$~meV is within the thermal activation uncertainty window of $\sim 1.6$~meV. It is likely that the specific details of the carrier redistribution processes become important within this energy window. We calculate statistical mean and variance of the zero-crossing energy of ECPL signal at different delays to estimate the mobility edge at $0.4 \pm 0.5$~meV above the exciton PL peak. Consistent with previous reports, this lies within the Stokes shift energy from the exciton absorption peak. We may note that the ECPL signal at energies of about 10~meV below the 1s hh-x PL peak, while positive at shorter delays ($< 400$~ps) tends to become negative at sufficiently long delays ($> 500$~ps). This may be due to very slow migration of low energy localized excitons to further lower energy states, in the time-scale longer than $\tau_r$.

\section{Summary}
To summarize, we have shown that the nonlinear time-resolved excitation correlation photoluminescence measurements can be conveniently performed with high energy resolution by numerically subtracting the spectra measured as a function of delay between two excitation pulses using a CCD-based spectrograph. Being a differential measurement, ECPL spectroscopy is very sensitive to various spectral features and to different carrier dynamics related to these spectral features. We use this measurement technique to (i) address the issue of coexistence of extended excitonic states under weak disorder with highly localized excitonic states due to strong disorder in the same semiconductor QW structure and (ii) to determine the location of the `mobility edge' which is the energy separating the QW-like extended and the QD-like localized excitonic states. Measurements on a high-quality GaAs MQWs sample showed that the dynamics of carriers above and below the heavy-hole exciton peak is different but complimentary to each other. The dynamics is governed by the carrier migration to lower energy, controlled by the carrier recombination dynamics and Pauli blocking effect. A change in sign of the ECPL signal as a function of energy is observed with the signal crossing zero within a narrow band of energy slightly above the center of the exciton emission line. This energy value is interpreted as the excitonic mobility edge. This demonstrates ECPL spectroscopy as a new and convenient method of experimentally locating mobility edge. Although the PL and ECPL spectra were measured with high energy resolution, determination of mobility edge suffers from an energy uncertainty of about 1~meV coming from the finite temperature of the experiment. 

\begin{acknowledgments}  
We thankfully acknowledge financial support from MHRD and DST, India.
\end{acknowledgments}

\end{document}